\documentclass[10pt,a4paper]{article}
\usepackage{lineno}
\usepackage{graphicx}
\usepackage{subfigure}
\usepackage{xspace}
\usepackage{color}
\usepackage{colortbl}
\usepackage{amsmath}
\usepackage{ifthen}
\usepackage{multirow}
\usepackage{multicol}
\usepackage{amssymb}
\usepackage{amsfonts}
\usepackage{upgreek}
\usepackage{hyperref}
\usepackage[all]{hypcap}
\usepackage{mciteplus}
\newboolean{articletitles}
\setboolean{articletitles}{true} 
\newboolean{uprightparticles}
\setboolean{uprightparticles}{false} 
\def\lhcb {LHCb\xspace}
\def\ux85 {UX85\xspace}

\def\babar  {BaBar\xspace}

\def\dzero  {D0\xspace}

\ifthenelse{\boolean{uprightparticles}}%
{

 \def\Pmu         {\ensuremath{\upmu}\xspace}

 \def\Ppi         {\ensuremath{\uppi}\xspace}

 \def\Ppsi        {\ensuremath{\uppsi}\xspace}

 \def\PDelta      {\ensuremath{\Delta}\xspace}                 
 \def\PXi      {\ensuremath{\Xi}\xspace}                 
 \def\PLambda      {\ensuremath{\Lambda}\xspace}                 
 \def\PSigma      {\ensuremath{\Sigma}\xspace}                 
 \def\POmega      {\ensuremath{\Omega}\xspace}                 
 \def\PUpsilon      {\ensuremath{\Upsilon}\xspace}                 
 
 \def\PB      {\ensuremath{\mathrm{B}}\xspace}                 
                  
 \def\PD      {\ensuremath{\mathrm{D}}\xspace}

 \def\PJ      {\ensuremath{\mathrm{J}}\xspace}                 
 \def\PK      {\ensuremath{\mathrm{K}}\xspace}

 \def\Pi      {\ensuremath{\mathrm{i}}\xspace}

 \def\Ps      {\ensuremath{\mathrm{s}}\xspace}

}
{

 \def\Pmu         {\ensuremath{\mu}\xspace}

 \def\Ppi         {\ensuremath{\pi}\xspace}

 \def\Ppsi        {\ensuremath{\psi}\xspace}                 
                  
 \mathchardef\PDelta="7101
 \mathchardef\PXi="7104
 \mathchardef\PLambda="7103
 \mathchardef\PSigma="7106
 \mathchardef\POmega="710A
 \mathchardef\PUpsilon="7107
                  
 \def\PB      {\ensuremath{B}\xspace}                 
                  
 \def\PD      {\ensuremath{D}\xspace}

 \def\PJ      {\ensuremath{J}\xspace}                 
 \def\PK      {\ensuremath{K}\xspace}

 \def\Pi      {\ensuremath{i}\xspace}

 \def\Ps      {\ensuremath{s}\xspace}

}

\def\mup        {\ensuremath{\Pmu^+}\xspace}
\def\mun        {\ensuremath{\Pmu^-}\xspace} 

\def\squark    {\ensuremath{\Ps}\xspace}

\def\pion  {\ensuremath{\Ppi}\xspace}

\def\pip   {\ensuremath{\pion^+}\xspace}
\def\pim   {\ensuremath{\pion^-}\xspace}

\def\pipm  {\ensuremath{\pion^\pm}\xspace}

\def\kaon  {\ensuremath{\PK}\xspace}
  \def\Kbar  {\kern 0.2em\overline{\kern -0.2em \PK}{}\xspace}

\def\Kz    {\ensuremath{\kaon^0}\xspace}
\def\Kzb   {\ensuremath{\Kbar^0}\xspace}
\def\KzKzb {\ensuremath{\Kz \kern -0.16em \Kzb}\xspace}
\def\Kp    {\ensuremath{\kaon^+}\xspace}
\def\Km    {\ensuremath{\kaon^-}\xspace}
\def\Kpm   {\ensuremath{\kaon^\pm}\xspace}

\def\KpKm  {\ensuremath{\Kp \kern -0.16em \Km}\xspace}

  \def\Dbar    {\kern 0.2em\overline{\kern -0.2em \PD}{}\xspace}
\def\D       {\ensuremath{\PD}\xspace}

\def\Dz      {\ensuremath{\D^0}\xspace}
\def\Dzb     {\ensuremath{\Dbar^0}\xspace}
\def\DzDzb   {\ensuremath{\Dz {\kern -0.16em \Dzb}}\xspace}
\def\Dp      {\ensuremath{\D^+}\xspace}
\def\Dm      {\ensuremath{\D^-}\xspace}

\def\DpDm    {\ensuremath{\Dp {\kern -0.16em \Dm}}\xspace}

\def\Dstarp  {\ensuremath{\D^{*+}}\xspace}
\def\Dstarm  {\ensuremath{\D^{*-}}\xspace}

\def\B       {\ensuremath{\PB}\xspace}
  \def\Bbar    {\kern 0.18em\overline{\kern -0.18em \PB}{}\xspace}

\def\Bz      {\ensuremath{\B^0}\xspace}

\def\Bu      {\ensuremath{\B^+}\xspace}
\def\Bub     {\ensuremath{\B^-}\xspace}
\def\Bp      {\ensuremath{\Bu}\xspace}
\def\Bm      {\ensuremath{\Bub}\xspace}
\def\Bpm     {\ensuremath{\B^\pm}\xspace}

\def\Bd      {\ensuremath{\B^0}\xspace}
\def\Bs      {\ensuremath{\B^0_\squark}\xspace}
\def\Bsb     {\ensuremath{\Bbar^0_\squark}\xspace}

\def\jpsi     {\ensuremath{{\PJ\mskip -3mu/\mskip -2mu\Ppsi\mskip 2mu}}\xspace}
\def\psitwos  {\ensuremath{\Ppsi{(2S)}}\xspace}

  \def\Y#1S{\ensuremath{\PUpsilon{(#1S)}}\xspace}

\def\to                 {\ensuremath{\rightarrow}\xspace}

\def\CP                {\ensuremath{C\!P}\xspace}

\def\AT#1     {\ensuremath{A_{\mathrm{T}}^{#1}}\xspace}           

\def\C#1      {\ensuremath{\mathcal{C}_{#1}}\xspace}                       
\def\Cp#1     {\ensuremath{\mathcal{C}_{#1}^{'}}\xspace}                    
\def\Ceff#1   {\ensuremath{\mathcal{C}_{#1}^{\mathrm{(eff)}}}\xspace}        
\def\Cpeff#1  {\ensuremath{\mathcal{C}_{#1}^{'\mathrm{(eff)}}}\xspace}       
\def\Ope#1    {\ensuremath{\mathcal{O}_{#1}}\xspace}                       
\def\Opep#1   {\ensuremath{\mathcal{O}_{#1}^{'}}\xspace}                    

\newcommand{\tev}{\ensuremath{\mathrm{\,Te\kern -0.1em V}}\xspace}
\newcommand{\gev}{\ensuremath{\mathrm{\,Ge\kern -0.1em V}}\xspace}
\newcommand{\mev}{\ensuremath{\mathrm{\,Me\kern -0.1em V}}\xspace}
\newcommand{\kev}{\ensuremath{\mathrm{\,ke\kern -0.1em V}}\xspace}
\newcommand{\ev}{\ensuremath{\mathrm{\,e\kern -0.1em V}}\xspace}
\newcommand{\gevc}{\ensuremath{{\mathrm{\,Ge\kern -0.1em V\!/}c}}\xspace}
\newcommand{\mevc}{\ensuremath{{\mathrm{\,Me\kern -0.1em V\!/}c}}\xspace}
\newcommand{\gevcc}{\ensuremath{{\mathrm{\,Ge\kern -0.1em V\!/}c^2}}\xspace}
\newcommand{\gevgevcccc}{\ensuremath{{\mathrm{\,Ge\kern -0.1em V^2\!/}c^4}}\xspace}
\newcommand{\mevcc}{\ensuremath{{\mathrm{\,Me\kern -0.1em V\!/}c^2}}\xspace}

\def\mm   {\ensuremath{\rm \,mm}\xspace}

\def\invfb   {\ensuremath{\mbox{\,fb}^{-1}}\xspace}

\def\mhz  {\ensuremath{{\rm \,MHz}}\xspace}
\def\khz  {\ensuremath{{\rm \,kHz}}\xspace}

\def\gsim{{~\raise.15em\hbox{$>$}\kern-.85em
          \lower.35em\hbox{$\sim$}~}\xspace}
\def\lsim{{~\raise.15em\hbox{$<$}\kern-.85em
          \lower.35em\hbox{$\sim$}~}\xspace}

\def\pt         {\mbox{$p_{\rm T}$}\xspace}

\def\mrad{\ensuremath{\rm \,mrad}\xspace}

\def\evtgen     {\mbox{\textsc{EvtGen}}\xspace}
\def\pythia     {\mbox{\textsc{Pythia}}\xspace}

\def\geant      {\mbox{\textsc{Geant4}}\xspace}

\def\tell1  {TELL1\xspace}
\def\ukl1   {UKL1\xspace}

\begin{document}

\begin{titlepage}
\pagenumbering{roman}
\vspace*{-1.5cm}
\centerline{\large EUROPEAN ORGANIZATION FOR NUCLEAR RESEARCH (CERN)}
\vspace*{1.5cm}
\hspace*{-0.5cm}
\begin{tabular*}{\linewidth}{lc@{\extracolsep{\fill}}r}
\vspace*{-1.4cm}\mbox{\!\!\!\includegraphics[width=.12\textwidth]{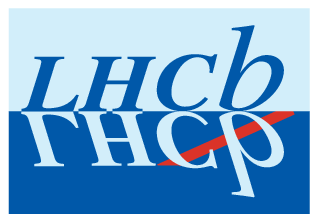}} & &
\\
 & & LHCB-PAPER-2011-024 \\
 & & CERN-PH-EP-2012-070 \\
 & & \today \\
 & & \\
\end{tabular*}
\vspace*{3.0cm}
{\bf\boldmath\Large
\begin{center}
Measurements of the branching fractions and $C\!P$ asymmetries\\ of $B^{\pm} \to J\!/\!\psi\, \pi^{\pm}$ and $B^{\pm} \to \psi(2S) \pi^{\pm}$ decays
\end{center}
}
\vspace*{2.0cm}
\begin{center}
The \lhcb collaboration
\footnote{Authors are listed on the following pages.}
\end{center}
\vspace*{2.0cm}
\begin{abstract}
\noindent
A study of $B^{\pm} \to J\!/\!\psi\, \pi^{\pm}$ and $B^{\pm} \to \psi(2S) \pi^{\pm}$ decays is performed with data corresponding to $0.37\,{\rm fb}^{-1}$ of proton-proton collisions at $\sqrt{s}=7\,\mathrm{Te\kern -0.1em V}$.
Their branching fractions are found to be
\begin{eqnarray*}
\mathcal{B}(B^{\pm} \to J\!/\!\psi\, \pi^{\pm}) &=& (3.88 \pm 0.11  \pm 0.15) \times 10^{-5}\ {\rm and}\\
\mathcal{B}(B^{\pm} \to \psi(2S) \pi^{\pm}) &=& (2.52 \pm 0.26  \pm 0.15) \times 10^{-5},
\end{eqnarray*}
where the first uncertainty is related to the statistical size of the sample and the second quantifies systematic effects.
The measured $C\!P$ asymmetries in these modes are
\begin{eqnarray*}
A_{CP}^{J\!/\!\psi\, \pi}    &=& 0.005 \pm 0.027 \pm 0.011\ {\rm and} \\
A_{CP}^{\psi(2S) \pi} &=& 0.048 \pm 0.090 \pm 0.011
\end{eqnarray*}
with no evidence of direct $C\!P$ violation seen.
\end{abstract}
\vspace*{2.0cm}
\begin{center}
{\it Submitted to Phys.\ Rev.\ X}
\end{center}
\vspace{2.0cm}
\end{titlepage}
\newpage
\setcounter{page}{2}
\onecolumn
\begin{center}
\begin{small}
{\bf \lhcb collaboration}\\

\begin{flushleft}
R.~Aaij$^{38}$, 
C.~Abellan~Beteta$^{33,n}$, 
B.~Adeva$^{34}$, 
M.~Adinolfi$^{43}$, 
C.~Adrover$^{6}$, 
A.~Affolder$^{49}$, 
Z.~Ajaltouni$^{5}$, 
J.~Albrecht$^{35}$, 
F.~Alessio$^{35}$, 
M.~Alexander$^{48}$
% ...

{\footnotesize \it
% ...
$ ^{5}$Clermont Universit\'{e}, Universit\'{e} Blaise Pascal, CNRS/IN2P3, LPC, Clermont-Ferrand, France\\
$ ^{6}$CPPM, Aix-Marseille Universit\'{e}, CNRS/IN2P3, Marseille, France\\
% ...
$ ^{33}$Universitat de Barcelona, Barcelona, Spain\\
$ ^{34}$Universidad de Santiago de Compostela, Santiago de Compostela, Spain\\
$ ^{35}$European Organization for Nuclear Research (CERN), Geneva, Switzerland\\
% ...
$ ^{38}$Nikhef National Institute for Subatomic Physics, Amsterdam, The Netherlands\\
% ...
$ ^{43}$H.H. Wills Physics Laboratory, University of Bristol, Bristol, United Kingdom\\
% ...
$ ^{48}$School of Physics and Astronomy, University of Glasgow, Glasgow, United Kingdom\\
$ ^{49}$Oliver Lodge Laboratory, University of Liverpool, Liverpool, United Kingdom\\
% ...
$ ^{n}$LIFAELS, La Salle, Universitat Ramon Llull, Barcelona, Spain\\
% ...
}
\bigskip
\end{flushleft}

\end{small}
\end{center}
\newpage
\twocolumn
\pagestyle{plain}
\setcounter{page}{1}
\pagenumbering{arabic}
\twocolumn

\noindent
The Cabibbo-suppressed decay $\Bp\to\psi\pip$, where $\psi$ represents either a $\jpsi$ or $\psitwos$, proceeds via a $b \to c\bar{c}d$ quark transition. 
Its branching fraction is expected to be about 5\% of the favoured $b \to c\bar{c}s$ mode, $\Bp\to\psi\Kp$ (charge conjugation is implied unless otherwise stated). 
The Standard Model predicts that for $b \to c\bar{c}s$ decays the tree and penguin contributions have the same weak phase and thus no direct \CP violation is expected in $\Bp\to\psi\Kp$.
For $\Bp\to\psi\pip$, the tree and penguin contributions have different phases and \CP asymmetries at the per mille level may occur~\cite{Dunietz:1993cg}.
An additional asymmetry may be generated, at the percent level, from long-distance rescattering, particularly from decays that have the same quark content ($\Dz\Dm$, $\Dstarm\Dz$, ...)~\cite{Soares:1994vi}. 
Any asymmetry larger than this would be of significant interest.\\

In this paper, the \CP asymmetries
\begin{equation}
A^{\psi \pi} = \frac{ \mathcal{B}(\Bm \to \psi \pi^{-}) - \mathcal{B}(\Bp \to \psi \pip ) } { \mathcal{B}(\Bm \to \psi \pi^{-}) + \mathcal{B}(\Bp \to \psi \pip)}
\label{eq:Acp}
\end{equation}
and charge-averaged ratios of branching fractions
\begin{equation}
R^{\psi} = \frac{\mathcal{B}(B^{\pm} \to \psi \pi^{\pm})} {\mathcal{B}(B^{\pm} \to \psi K^{\pm})}
\label{eq:R}
\end{equation}
are measured with the $\psi$ reconstructed in the $\mup\mun$ final state. 
From the latter, $\mathcal{B}(B^{\pm} \to \psi \pi^{\pm})$ may be deduced using the established $B^{\pm} \to \psi K^{\pm}$ branching fractions~\cite{Nakamura:2010zzi}.
The \CP asymmetry for $\Bp\to\psitwos\Kp$ is also reported. 
$\Bp\to\jpsi\Kp$ acts as a control mode in the asymmetry analysis because it is well measured and no \CP violation is observed~\cite{Nakamura:2010zzi}.
Previous measurements of the $\Bp\to\jpsi\pip$ branching fractions and \CP asymmetries~\cite{Aubert:2004pra,Abazov:2008gs} have an accuracy of about 10\%.
The $\Bp \to \psitwos h^{+}\ (h = K, \pi)$ system is less precisely known due to a factor ten lower branching fraction to the $h\mu\mu$ final state.
The world average for $A^{\psitwos K}$ is $-0.025\pm0.024$~\cite{Nakamura:2010zzi} and there has been one measurement of $A^{\psitwos \pi}=0.022\pm0.086$~\cite{Bhardwaj:2008ee}.\\

The \lhcb experiment~\cite{Alves:2008zz} takes advantage of the high $b\bar{b}$ and $c\bar{c}$ cross sections at the Large Hadron Collider to record unprecedented samples of heavy hadron decays.
It instruments the pseudorapidity range $2<\eta <5$ of the proton-proton ($pp$) collisions with a dipole magnet and a tracking system which achieves a momentum resolution of $0.4-0.6\%$ in the range $5-100$~\gevc. 
The dipole magnet can be operated in either polarity and this feature is used to reduce systematic effects due to detector asymmetries. In the sample analysed here, 55\% of data was taken with one polarity, 45\% with the other.\\

The $pp$ collisions take place inside a silicon-strip vertex detector which has active material $8 \mm$ from the beam line. 
It provides measurements of track impact parameters (IP) with respect to primary collision vertices (PV) and precise reconstruction of secondary \Bp vertices. 
Downstream muon stations identify muons by their penetration through layers of iron shielding. 
Charged particle identification (PID) is realised using ring-imaging Cherenkov (RICH) detectors with three radiators: aerogel, ${\rm C}_4{\rm F}_{10}$ and ${\rm CF}_{4}$.
Events with a high transverse energy cluster in calorimeters or a high transverse momentum (\pt) muon activate a hardware trigger. 
About $1 \mhz$ of such events are passed to a software-implemented high level trigger, which retains about $3 \khz$.\\

The analysis is performed using $0.37~\invfb$ of data recorded by LHCb in the first half of 2011.
The decay chain $\Bp\to\psi h^{+},\ \psi\to\mup\mun$ is reconstructed from good quality tracks which have a track-fit $\chi^{2}$ per degree of freedom $< 5$. 
The muons are required to have momentum, $p > 3 \gevc$, and $\pt > 0.5 \gevc$. 
Selected hadrons have $p > 5 \gevc$ and $\pt > 1 \gevc$. 
The two muon candidates are used to form a $\psi$ resonance with vertex-fit $\chi^{2}<10$.
The dimuon invariant mass is required to be within $^{+30}_{-40}$~\mevcc of the nominal $\psi$ mass~\cite{Nakamura:2010zzi}; the asymmetric limits allow for a radiative tail.\\

The reconstructed \Bp candidate vertex is required to be of good quality with a vertex-fit $\chi^{2}<10$.
It is ensured to originate from a PV by requiring $\chi_{\rm IP}^{2} < 25$ where the $\chi^2$ considers the uncertainty on track IP and the PV position.
In addition, the angle between the \Bp momentum vector and its direction of flight from the PV must be $<32~(10)$~\mrad for $\psitwos h^{+}$ ($\jpsi h^{+}$). 
Furthermore, neither the muons nor the hadron track may point back to any primary vertex with $\chi_{\rm IP}^2$ $<4$.
It is required that the hardware trigger accepted a muon from the \Bp candidate or by activity in the rest of the event.
Hardware-trigger decisions based on the hadron are neglected to remove dependence on the correct emulation of the calorimeter's response to pions and kaons.\\

The \Bp candidates are refitted~\cite{2005NIMPA.552} requiring all three tracks to originate from the same point in space and the $\psi$ candidates to have their nominal mass~\cite{Nakamura:2010zzi}. 
Candidates for which one muon gives rise to two tracks in the reconstruction, one of which is then assumed to be the hadron, form an artificial peaking background in the $\psitwos h^+$ analysis.
These candidates peak in the invariant mass distribution of the same-sign muon-pion combination at $m_{\mu\pi}\sim245$~\mevcc, i.e. the sum of the muon and pion rest masses.
Requiring $m_{\mu\pi}> 300$~\mevcc removes this background. 
In 2\% of events two \Bp candidates are found.
If they decay within 2 \mm of each other the candidate with the poorest quality vertex is removed; otherwise both are kept.\\

\begin{figure}[htb]
\begin{center}
\includegraphics[width=0.99\columnwidth]{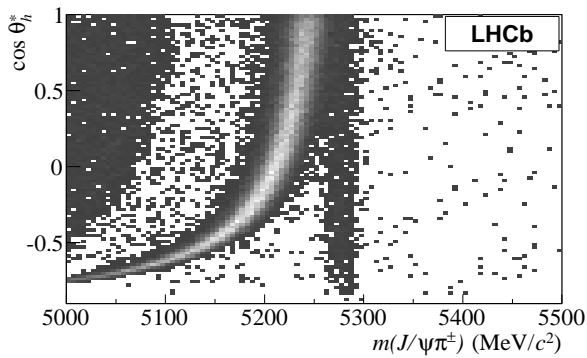}
\caption{
Distribution of $\cos(\theta^*_h)$ versus the invariant mass of $\Bp \to \jpsi \pip$ candidates. 
The curved structure contains misidentified $\Bp \to \jpsi \Kp$ decays which separate from the $\Bp \to \jpsi \pip$ vertical band for $\cos(\theta^*_h)<0$.
The partially reconstructed background, $B\to \jpsi K\pi$ enters top left.
\label{fig:hadron_decay_angle}}
\end{center}
\end{figure}

When selecting $\jpsi h^{+}$ candidates, a requirement is made on the decay angle of the charged hadron as measured in the rest frame of the \Bp with respect to the \Bp trajectory in the laboratory frame, $\cos(\theta^*_h)<0$. 
This requires the hadron to have flown counter to the trajectory of the \Bp candidate, hence lowering its average momentum in the laboratory frame.
At lower momentum, the pion-kaon mass difference provides sufficient separation in the \Bp invariant mass distribution, as shown in Fig.~\ref{fig:hadron_decay_angle}. 
In the $\Bp \to \psitwos h^{+}$ analysis, the average momentum of the hadrons is lower, so such a cut is unecessary to separate the two modes.\\

Particle identification information is quantified as differences between the logarithm of likelihoods, $\ln\mathcal{L}_h$, under five mass hypotheses, $h \in \{\pi,\ K,\ p,\ e,\ \mu\}$.
Separation of $\psi\pip$ candidates from $\psi\Kp$ is ensured by requiring that the hadron track satisfies $\ln\mathcal{L}_{K}-\ln\mathcal{L}_{\pi}= {\rm DLL}_{K\pi} <6$.
This value is chosen to ensure that most ($\sim95\%$) $\Bp \to \psi \pip$ decays are reconstructed as such.
These events form the ``pion-like" sample, as opposed to the kaon-like events satisfying ${\rm DLL}_{K\pi} > 6$ that are reconstructed under the $\psi\Kp$ hypothesis.\\

\begin{figure*}[htbp]
\begin{center}
\includegraphics*[width=0.95 \textwidth]{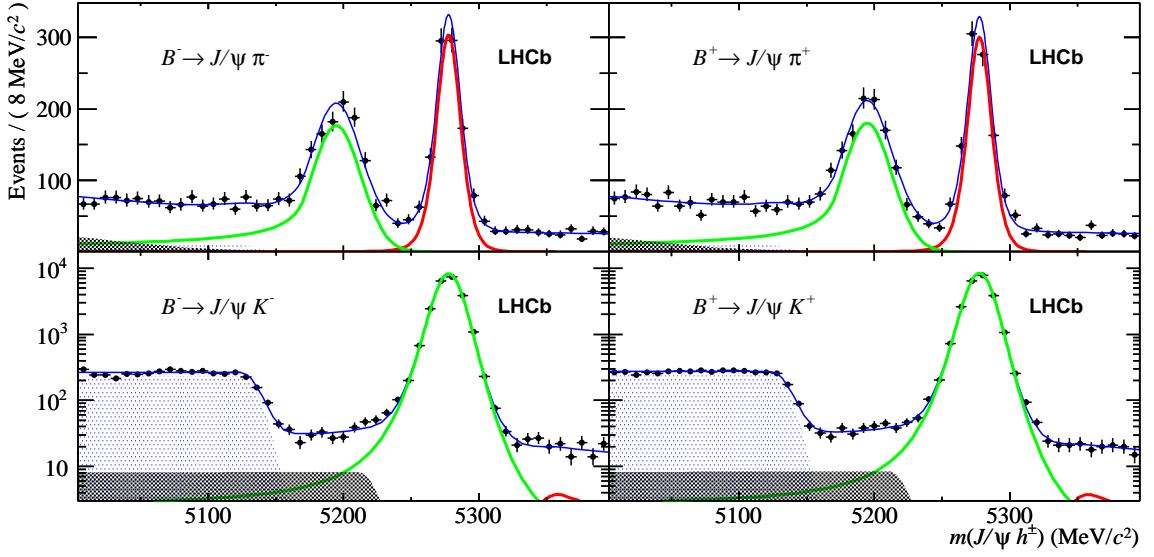}
\end{center}
\caption{
Distributions of $\Bpm \to \jpsi h^{\pm}$ invariant mass, overlain by the total fitted PDF (thin line).
Pion-like events, with DLL$_{K\pi}<6$ are reconstructed as $\jpsi\pipm$ and enter in the top plots.
All other events are reconstructed as $\jpsi\Kpm$ and are shown in the bottom plots on a logarithmic scale.
\Bm decays are shown on the left, \Bp on the right.
The dark [red] curve shows the $\Bpm\to\jpsi\pipm$ component, the light [green] curve represents $\Bpm\to\jpsi\Kpm$.
The partially reconstructed contributions are shaded.
In the lower plots these are visualised with a dark (light) shade for \Bs(\Bu or \Bd) decays. 
In the top plots the shaded component are contributions from $\B\to\jpsi\Kpm\pi$ (dark) and $\B\to\jpsi\pipm\pi$ (light).
\label{fig:fit}}
\end{figure*}

\begin{figure*}[htbp]
\begin{center}
\includegraphics*[width=0.95 \textwidth]{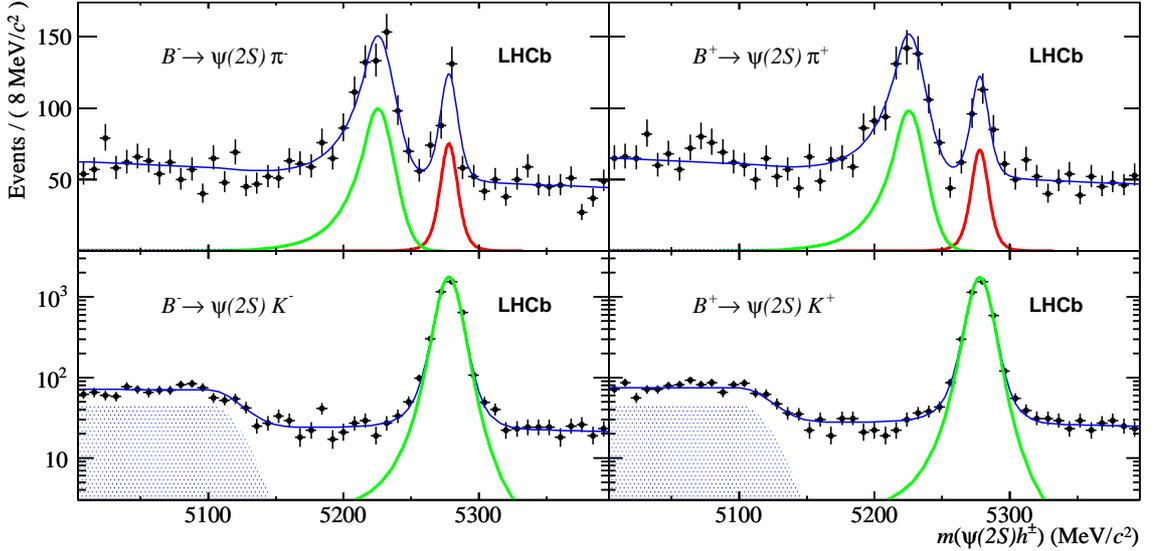}
\end{center}
\caption{Distributions of $\Bpm \to \psitwos h^{\pm}$ invariant mass. See the caption of Fig.~\ref{fig:fit} for details. 
The partially reconstructed background in the pion-like sample is present but negligible yields are found.
\label{fig:fit2s}}
\end{figure*}

The selected data are partitioned by magnet polarity, charge and $\mathrm{DLL}_{K\pi}$ of the hadron track.
By keeping the two magnet polarity samples separate, residual detection asymmetries between the left and right sides of the detector can be evaluated and hence factor out. 
Event yields are extracted by performing an unbinned, maximum-likelihood fit simultaneously to the eight distributions of \B invariant mass in the range $5000<m_{B}<5780~\mevcc$~\cite{2003physics6116V}.
Figure~\ref{fig:fit} shows this fit to the data for $\Bp \to \jpsi h^{+}$, summed over magnet polarity. The $\Bp \to \psitwos h^{+}$ data is shown in Fig.~\ref{fig:fit2s}.\\

The probability density function (PDF) used to describe these distributions has several components.
The correctly reconstructed, $\Bp \to \psi h^{+}$ events are modelled by the function,
\begin{equation}
f(x) \propto \exp\left(\frac{-(x-\mu)^2}{2\sigma^2+(x-\mu)^2\alpha_{L,R}}\right)
\end{equation}
which describes an asymmetric peak of mean $\mu$ and width $\sigma$, and where $\alpha_L(x<\mu)$ and $\alpha_R(x>\mu)$ parameterise the tails.
The mean is required to be the same for $\psi\Kp$ and $\psi\pip$ though it can vary across the four charge$\times$polarity subsamples to account for different misalignment effects.
Table~\ref{table:shape_results} shows the fitted values of the common tail parameters and the widths of the $\Bp\to\psi h^+$ peaks averaged over the subsamples.
\begin{table}[h]
\centering
\caption{Signal shape parameters from the $\Bpm\to\psi h^{\pm}$ fits.
\label{table:shape_results}}
\vspace{2mm}

\begin{tabular}{ l l | c  c }
& & \jpsi & \psitwos \\ 
\hline
$\sigma_{\psi K}$ &(\mevcc)  & $7.84\pm0.04$ & $6.02\pm0.08$ \\ 
$\sigma_{\psi\pi}$ &(\mevcc)  & $8.58\pm0.27$ & $6.12\pm0.75$ \\
$\alpha_{\rm L}$& & $0.12\pm0.03$ & $0.14\pm0.01$ \\
$\alpha_{\rm R}$& & $0.10\pm0.03$ & $0.13\pm0.01$ \\
\end{tabular}

\end{table}

The misidentified $\psi \Kp$ events form a displaced peaking structure to the left of the $\psi \pip$ signal and tapers to lower mass. 
This is modelled by a Crystal Ball function~\cite{Skwarnicki:1986xj} which is found to be a suitable effective PDF.
Its yield is added to that of the correctly identified events to calculate the total number of $\psi \Kp$ events.\\

The PDF modelling the small component of $\psi \pip$ decays with DLL$_{K\pi}>6$ is fixed entirely from simulation. 
It contributes negligibly to the total likelihood so the yield must be fixed with respect to that of correctly identified $\psi \pip$ events.
The efficiency of the PID cut is estimated using samples of pions and kaons from $\Dz\to\Kp\pim$ decays which are selected with high purity without using PID information.
These calibration events are reweighted in bins of momentum to match the momentum distribution of the large $\jpsi\Kp$ and $\psitwos\Kp$ samples.
By this technique, the following efficiencies are deduced for ${\rm DLL}_{K\pi}<6$:
$\epsilon_{\jpsi\pi}=(95.8\pm1.0)\%$;
$\epsilon_{\psitwos\pi}=(96.6\pm1.0)\%$.
The errors, estimated from simulation, account for imperfections in the reweighting and the difference of the signal \Kp and \pip momenta.\\

Partially reconstructed decays populate the region below the \Bp mass. 
$B^{+/0}\to\psi\Kp\pi$ decays, where the pion is missed, are modelled in the kaon-like sample by a flat PDF with a Gaussian edge.
A small $\Bsb\to\psi\Kp\pim$ component is needed to achieve a stable fit. 
It is modelled with the same shape as the partially reconstructed $B^{+/0}$ decays except shifted in mass by the \Bs-\Bz mass difference, $+87$~\mevcc. 
In the pion-like sample, $\psi\pip\pi$ backgrounds are assumed to enter with the same PDF, and same proportion relative to the signal, as the $\psi\Kp\pi$ background in the kaon-like sample.
A component of misidentified $B^{+/0}\to\jpsi\Kp\pi$ is also included with a fixed shape estimated from the data.
Lastly, a linear polynomial with a negative gradient is used to approximate the combinatorial background. 
The slope of this component of the pion-like and kaon-like backgrounds can differ.\\

\begin{table}[h]
\centering
\caption{Raw fitted yields. The labels `D' and `U' refer to the two polarities of the \lhcb dipole.}
\vspace{2mm}
\begin{tabular}{ c  c  c | c  c }
& & & \Bm & \Bp  \\ 
\hline&&\\[-2.5ex]
\multirow{4}{*}{\jpsi}    & \multirow{2}{*}{$\pi$} & D & $\phantom{00\,}528\pm\phantom{0}27$ & $\phantom{00\,}518\pm\phantom{0}27$ \\
                          &                        & U & $\phantom{00\,}421\pm\phantom{0}23$ & $\phantom{00\,}428\pm\phantom{0}23$ \\ 
                          & \multirow{2}{*}{$K$}   & D & $\phantom{}13\,363\pm\phantom{}180$ & $\phantom{}13\,466\pm\phantom{}181$ \\
                          &                        & U & $\phantom{}10\,666\pm\phantom{}148$ & $\phantom{}11\,120\pm\phantom{}155$ \\
\hline&&\\[-2.5ex]
\multirow{4}{*}{\psitwos} & \multirow{2}{*}{$\pi$} & D & $\phantom{00\,0}94\pm\phantom{0}16$ & $\phantom{00\,0}93\pm\phantom{0}16$ \\
                          &                        & U & $\phantom{00\,0}82\pm\phantom{0}15$ & $\phantom{00\,0}70\pm\phantom{0}13$ \\
                          & \multirow{2}{*}{$K$}   & D & $\phantom{0}2\,331\pm\phantom{0}88$ & $\phantom{0}2\,463\pm\phantom{0}93$ \\
                          &                        & U & $\phantom{0}2\,026\pm\phantom{0}78$ & $\phantom{0}1\,836\pm\phantom{0}71$ \\
\end{tabular}
\label{table:raw_fit_results}
\end{table}

The stability of the fit is tested with a large sample of pseudo-experiments.
Pull distributions from these tests are consistent with being normally distributed, demonstrating that the fit is stable under statistical variations. 
The yields obtained from the signal extraction fit are shown in Table~\ref{table:raw_fit_results}.\\

The observables, defined in Eqs.~\ref{eq:Acp} and \ref{eq:R} are calculated by the fit, then modified by a set of corrections taken from simulation.
The acceptances of $\psi\pip$ and $\psi\Kp$ events in the detector are computed using \pythia~\cite{Sjostrand:2006za} to generate the primary collision and \evtgen~\cite{Lange:2001uf} to model the \Bp decay.
The efficiency of reconstructing and selecting $\psi\pip$ and $\psi\Kp$ decays is estimated with a bespoke simulation of LHCb based on \geant~\cite{Agostinelli:2002hh}.
It models the interaction of muons and the two hadron species with the detector material. 
The total correction $\epsilon^{\psi K}\!/\epsilon^{\psi\pi}$ is $0.985 \pm 0.012$ and $1.007 \pm 0.021$ for $R^{\jpsi}$ and $R^{\psitwos}$ respectively.\\

\CP asymmetries are extracted from the observed charge asymmetries $(A_{\rm Raw})$ by taking account of instrumentation effects. 
The interaction asymmetry of kaons, $A_{\rm Det}^{K}$ is expected to be non-zero, especially for low-momentum particles. 
This asymmetry, measured at LHCb using a sample of $\Dstarp\to\Dz\pip,\ \Dz\to\Kp\pim$ decays, is $-0.010 \pm 0.002$ if the pion asymmetry is zero~\cite{Aaij:2012qe}.
The null-asymmetry assumption for pions has been verified at \lhcb to an accuracy of $0.25$\%~\cite{LHCb-PAPER-2012-009}.
These results are used with enlarged uncertainties ($0.004$, for both kaons and pions) to account for the different momentum spectra of this sample and those used in the previous analyses. \\

In summary, the \CP asymmetry is defined as
\begin{equation}
A^{\psi h} = A_{\rm Raw}^{\psi h} - A_{\rm Prod} - A_{\rm Det}^{h},
\label{eq:acp}
\end{equation}
where the production asymmetry, $A_{\rm Prod}$, describes the different rates with which \Bm and \Bp hadronise out of the $pp$ collisions.
The observed, raw charge asymmetry in $\Bp\to\jpsi\Kp$ is $-0.012\pm0.004$. 
Using Eq.~\ref{eq:acp} with the established \CP asymmetry, $A^{\jpsi K}=0.001\pm0.007$~\cite{Nakamura:2010zzi}, $A_{\rm Prod}$ is estimated to be $-0.003 \pm 0.009$. 
This is applied as a correction to the other modes reported here.\\

\begin{table*}[htbp]
\centering
\caption{Summary of systematic uncertainties. The statistical fit errors are included for comparison.
\label{table:systematic_summary}}
\vspace{2mm}
\begin{tabular}{  l c  c  c  c  c  c  c  c  c |}
     &&  $R^{\jpsi} (\times10^{-2})$ & $A^{\jpsi \pi}$ && $R^{\psitwos} (\times 10^{-2})$ & $A^{\psitwos \pi}$ & $A^{\psitwos K}$ \\ 
\hline
Simulation uncertainty && $0.045$ & - && $0.088$ & - & - \\ 
PID efficiencies          && $0.043$ & - && $0.052$ & - & - \\
$A^{\jpsi K}$ (PDG~\cite{Nakamura:2010zzi}) && - & $0.0070$ && - & $0.0070$ & $0.0070$ \\
$A_{\rm Raw}^{\jpsi K}$ statistical error  && - & $0.0046$ && - & $0.0046$ & $0.0046$ \\
Detection asymmetries && - & $0.0056$ && - & $0.0056$ & - \\ 
Relative trigger efficiency &&                    $0.020$ & $0.0031$ && $0.050$ & $0.0036$ & $0.0003$ \\
Fixed fit parameters  && $0.005$ & $0.0006$ && $0.017$ & $0.0013$ & $0.0001$ \\
\hline
Sum in quadrature (syst.) && $0.065$ & $0.0106$ && $0.115$ & $0.0108$ & $0.0084$ \\ 
Fit error (stat.) && $0.110$ & $0.0268$ && $0.404$ & $0.0901$ & $0.0136$ \\ 
\end{tabular}
\end{table*}


The different contributions to the systematic uncertainties are summarised in Table~\ref{table:systematic_summary}.
They are assessed by modifying the final selection, or altering fixed parameters and rerunning the signal yield fit.
The maximum variation of each observable is taken as their systematic uncertainty.\\

The largest uncertainty is due to the use of simulation to estimate the acceptance and selection efficiencies.
It accounts for any bias due to imperfect modelling of the detector and its relative response to pions and kaons. 
Another important contribution arises from the loose trigger criteria that are employed. 
This uncertainty is estimated from the shift in the central values after rerunning the fit using only those events where the muons passed the software trigger.
The use of the PID calibration to estimate the efficiency for pions to the DLL$_{K\pi}<6$ selection also contributes a significant systematic uncertainty.\\

The measurements of $A^{\psi \pi}$ depend on the estimation of $A_{\rm Prod}$ from the $\Bp\to\jpsi\Kp$ channel. 
The uncertainty on $A_{\rm Prod}$ is determined by the statistical error of $A_{\rm Raw}^{\jpsi K}$ in the fit, 
the uncertainty on the world average of $A^{\jpsi K}$ and the estimation of $A_{\rm Det}^{h}$.
These effects are kept separate in the table where it is seen that the uncertainty on the nominal value of $A^{\jpsi K}$ dominates.
Finally, it is noted that the detector asymmetries cancel for $A^{\psitwos K}$ and a lower systematic uncertainty can be reported.\\

The measured ratios of branching fractions are
\begin{eqnarray*}
R^{\jpsi}             &=& (3.83 \pm 0.11  \pm 0.07) \times 10^{-2}  \\
R^{\psitwos}          &=& (3.95 \pm 0.40  \pm 0.12) \times 10^{-2},
\end{eqnarray*}
where the first uncertainty is statistical and the second systematic. 
$R^{\psitwos}$ is compatible with the one existing measurement, $(3.99 \pm 0.36 \pm 0.17) \times 10^{-2}$~\cite{Bhardwaj:2008ee}. 
The measurement of $R^{\jpsi}$ is $3.2 \sigma$ lower than the current world average, $(5.2 \pm  0.4) \times 10^{-2}$~\cite{Nakamura:2010zzi}.
Using the established measurements of the Cabibbo-favoured branching fractions~\cite{Nakamura:2010zzi}, we deduce 
\begin{eqnarray*}
\mathcal{B}(B^{\pm} \to \jpsi \pi^{\pm}) &\!\!\!\!=\!\!\!\!& (3.88 \pm 0.11  \pm 0.15) \times 10^{-5}\\
\mathcal{B}(B^{\pm} \to \psitwos \pi^{\pm}) &\!\!\!\!=\!\!\!\!& (2.52 \pm 0.26  \pm 0.15) \times 10^{-5},
\end{eqnarray*}
where the systematic uncertainties are summed in quadrature.
The measured \CP asymmetries,
\begin{eqnarray*}
A_{CP}^{\jpsi \pi}    &=& 0.005 \pm 0.027 \pm 0.011 \\
A_{CP}^{\psitwos \pi} &=& 0.048 \pm 0.090 \pm 0.011 \\
A_{CP}^{\psitwos K}   &=& 0.024 \pm 0.014 \pm 0.008 ,  
\end{eqnarray*}
have comparable or better precision than previous results, and no evidence of direct \CP violation is seen. 
\section*{Acknowledgements}
\noindent We express our gratitude to our colleagues in the CERN accelerator
departments for the excellent performance of the LHC. We thank the
technical and administrative staff at CERN and at the LHCb institutes,
and acknowledge support from the National Agencies: CAPES, CNPq,
FAPERJ and FINEP (Brazil); CERN; NSFC (China); CNRS/IN2P3 (France);
BMBF, DFG, HGF and MPG (Germany); SFI (Ireland); INFN (Italy); FOM and
NWO (The Netherlands); SCSR (Poland); ANCS (Romania); MinES of Russia and
Rosatom (Russia); MICINN, XuntaGal and GENCAT (Spain); SNSF and SER
(Switzerland); NAS Ukraine (Ukraine); STFC (United Kingdom); NSF
(USA). We also acknowledge the support received from the ERC under FP7
and the Region Auvergne.
\onecolumn
\ifx\mcitethebibliography\mciteundefinedmacro
\PackageError{LHCb.bst}{mciteplus.sty has not been loaded}
{This bibstyle requires the use of the mciteplus package.}\fi
\providecommand{\href}[2]{#2}

\end{document}